# BAYESIAN SOURCE SEPARATION APPLIED TO IDENTIFYING COMPLEX ORGANIC MOLECULES IN SPACE

*Kevin H. Knuth [A], Man Kit Tse [A], Joshua Choinsky [A], Haley A. Maunu [A], Duane F. Carbon [B]*

A. University at Albany, Department of Physics, Albany NY USA
B. NASA Ames Research Center, NASA Advanced Supercomputing Division, Moffett Field, CA, USA

**ABSTRACT**

Emission from a class of benzene-based molecules known as Polycyclic Aromatic Hydrocarbons (PAHs) dominates the infrared spectrum of star-forming regions. The observed emission appears to arise from the combined emission of numerous PAH species, each with its unique spectrum. Linear superposition of the PAH spectra identifies this problem as a source separation problem.

It is, however, of a formidable class of source separation problems given that different PAH sources potentially number in the hundreds, even thousands, and there is only one measured spectral signal for a given astrophysical site. Fortunately, the source spectra of the PAHs are known, but the signal is also contaminated by other spectral sources. We describe our ongoing work in developing Bayesian source separation techniques relying on nested sampling in conjunction with an ON/OFF mechanism enabling simultaneous estimation of the probability that a particular PAH species is present and its contribution to the spectrum.

## 1. INTRODUCTION

The infrared spectrum of star-forming regions (Figure 1) is dominated by emission from a class of benzene-based molecules known as Polycyclic Aromatic Hydrocarbons (PAHs). The presence of this emission is a principal indicator of star formation not only in our own Milky Way galaxy but also in distant galaxies [1]. The observed emission appears to arise from the combined emission of numerous PAH molecular species, both neutral and ionized, each with its unique spectrum. Strong evidence suggests that the concentrations of individual PAH species, degrees of ionization, and temperatures vary significantly between astrophysical sites. Unraveling these variations is crucial to a deeper understanding of star-forming regions in the universe. To date, despite attempts to fit data by hand [2], no one has successfully unscrambled the observed PAH emission of any astrophysical source into its individual

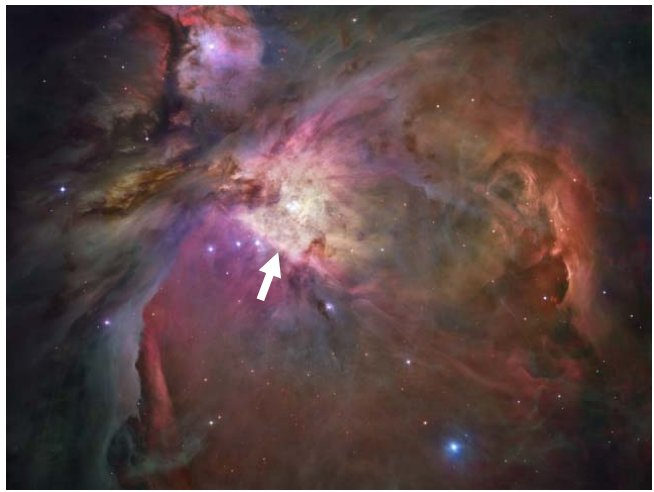

Figure 1. The Orion Nebula is an active star-forming region. The arrow indicates the Orion Bar where PAHs in the nebula are being excited by ultraviolet light from young stars. (HST/ACS Mosaic from STScI)

PAH constituents. Efforts have been defeated by the complexity of the observed PAH spectra and the very large number of potential PAH emitters.

Linear superposition of the various PAH species accompanied by additional sources identifies this problem as a source separation problem. It is, however, of a formidable class of source separation problems given that different PAH sources are potentially in the hundreds, even thousands, and there is only one measured spectral signal for a given astrophysical site. Fortunately, the source spectra of the PAHs are known, but the signal is also contaminated by other sources, such as dust radiation and atomic emission that must be simultaneously characterized to effectively isolate the PAHs.

Our ongoing work in developing informed Bayesian source separation techniques relies on incorporating three distinct types of source spectra models: a dictionary of over one thousand atomic and molecular PAH spectra in various states of ionization, Planck blackbody radiation parameterized by temperature to describe dust radiation, and a "non-parametric" mixture of Gaussians to describe poorly-understood spectral sources. These models are

Supported by NASA AISRP NNH05ZDA001N
http://aaaprod.gsfc.nasa.gov/aisrp/public/ProjectDetail.cfm?selectedProject=177

simultaneously estimated by a relatively new Markov chain Monte Carlo (MCMC) technique called Nested Sampling and by relying on an ON/OFF mechanism that enables us to simultaneously estimate the probability that a particular PAH species is present in addition to an estimation of its contribution to the spectrum.

In this paper we present a concise description of our initial investigations into this challenging problem.

## 2. PAHS AND THE INTERSTELLAR MEDIUM

A single PAH molecule is an assembly of hexagonally-shaped carbon rings of the simplest aromatic molecule, benzene, $C_6H_6$. The connected carbon rings of a PAH are laid out in flat structures resembling sections of chicken wire. Figure 2 shows two examples of PAHs along with their distinctive infra-red spectra, which arise from the vibrational states unique to their molecular structure. PAHs in the Interstellar Medium (ISM) are believed to range in size from a few 10s of Carbon atoms to hundreds of Carbon atoms [2]. In addition, they are thought to be present in neutral and ionic forms, and with substitutions of Deuterium and Nitrogen for some of the Hydrogen and Carbon atoms, respectively. Ionization and element substitution alter PAH wavelengths and relative intensities, often dramatically. While all PAHs have emission bands near 3.3, 6.2, 7.7, 8.6, 11.2, and 15-20 microns, they do not have identical spectra (see Figure 2). Our collaborators at NASA Ames Research Center have compiled a library of over a thousand PAH spectra, which we will be using as a template for the PAH sources. At this stage, we have investigated estimation of approximately 100 PAHs in simulations using a subset of this data. Our modeling technique attributes to each PAH species two parameters. The first is a binary parameter, which serves to turn the PAH ON and to turn it OFF. The second is a real parameter that serves to quantify the contribution of an ON PAH to the measured spectrum. This strategy enables us to not only quantify the probability with which we believe a particular PAH is present, but also the degree to which it is present. Atomic and ionic emissions are also present and must be modeled as well using their spectral templates from the database. Whether our algorithms or quality of results can be made to scale to a thousand PAHs is yet unknown, but that is the nature of this interesting problem.

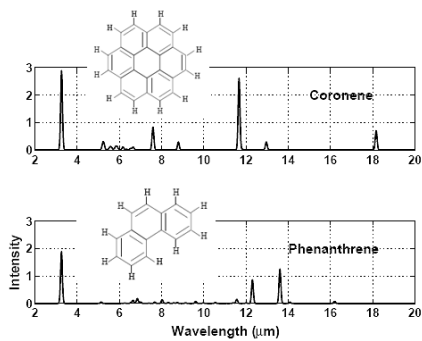

Figure 2. An example of two PAHs and their spectra.

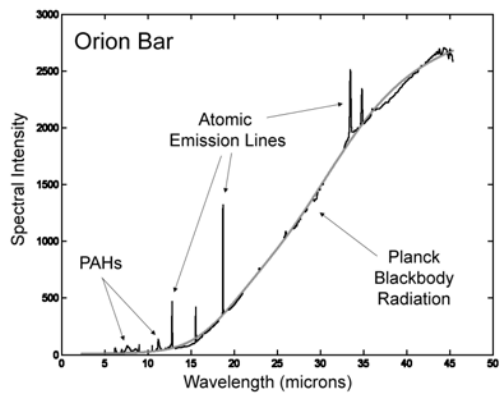

Figure 3. An infra-red spectrum taken from the Orion Bar of the Orion nebula. This is a dusty region illuminated by the ultraviolet light of young stars. The Planck blackbody radiation comprises the rising intensity at longer wavelengths. The tall sharp emission lines are from atoms and ions, and the small features between 2 and 15 microns are PAHs. The smooth gray curve is the estimated background describes in the Results section.

The measured spectrum (Figure 3) is complicated by the fact that the dust in the astrophysical site is thermally radiating. In fact, there may be several dust radiators along the line-of-sight, each at a distinct temperature (Figure 4). We assume, for this initial investigation, that the dust radiation can be modeled with multiple blackbody Planck functions of unknown intensity and temperature. This is a crude approximation but will serve our purpose here. Finally, there are broad spectral features that have not yet been attributed to any known source. To handle these unknown source components, we utilize a mixture of Gaussians with appropriately-defined cut-offs for both the mean and standard deviation.

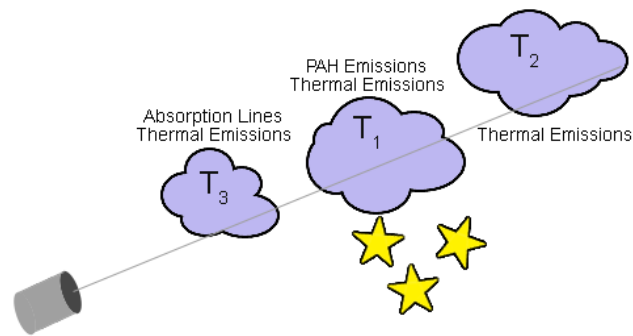

Figure 4. Ultraviolet light from young stars excite infrared PAH emissions a nearby interstellar cloud at temperature $T_1$. Since the nebulae are optically thin, the telescope also detects thermal radiation from clouds along the line-of-sight at temperatures $T_2$ and $T_3$, as well as possible absorption lines.

## 3. THE SPECTRUM MODEL

Blind techniques are not always useful in complex situations like these where much is known about the physics of the source signal generation and propagation. Higher-order models relying on physically-motivated parameterized functions are required, and by adopting such models, one can introduce more sophisticated likelihood and prior probabilities. We call this approach Informed Source Separation [3]. In this problem, we have linear mixing and the basic source separation equation holds:

$$F(\lambda) = \sum_{i=1}^{N} c_i s_i(\lambda) + \phi(\lambda), \quad (1)$$

where there are $N$ sources, each contributing a portion $c_i$ of their characteristic spectrum $s_i(\lambda)$ to the total measured flux $F(\lambda)$, which also suffers from a noise contribution $\phi(\lambda)$. Given our knowledge about the astrophysics, the summation can be rewritten as

$$\sum_{p=1}^{P} c_p PAH_p(\lambda) + \sum_{k=1}^{K} A_k Planck(\lambda;T_k) + \sum_{g=1}^{G} A_g N(\lambda;\bar{\lambda}_g,\sigma_g) \quad (2)$$

where $PAH_p$ is a $p$-indexed PAH spectrum from the dictionary, $N$ is a Gaussian. The function $Planck$ is

$$Planck(\lambda;T_k) = \sqrt{\frac{\lambda_{max}}{\lambda}} \frac{\exp(hc/\lambda_{max}kT)-1}{\exp(hc/\lambda kT)-1} \quad (3)$$

where $h$ is Planck's constant, $c$ is the speed of light, $k$ is Boltzmann's constant, $T$ is the temperature of the cloud, and $\lambda_{max}$ is the wavelength where the blackbody spectral energy peaks

$$\lambda_{max} = hc/4.965\,kT. \quad (4)$$

## 4. BAYESIAN METHODOLOGY

Although we possess a great deal of prior information regarding the interstellar environment, the individual spectral sources, and the details of the radiation processes, this is clearly a complex and challenging inference problem. To solve this problem while relying on our prior knowledge, we utilize the Bayesian methodology, which enables us to use recorded data to update our prior model. Bayes Theorem states that

$$P(M|D,I) = P(M|I)\frac{P(D|M,I)}{P(D|I)}, \quad (5)$$

where $M$ represents the set of model parameters, $D$ represents our data in the form of a recorded spectrum, and $I$ represents our prior information. The term $P(M|I)$ is called the prior probability as it represents the probability of the model parameter values given our prior information before we encounter any data. The likelihood $P(D|M,I)$ describes the probability that the hypothesized model $M$ could have led to the observed data. The posterior probability, or posterior for short, $P(M|D,I)$ describes the probability of the model parameters given both our prior information and the data. The denominator $P(D|I)$ describes the evidence that that data provides the model. In most cases, this is determined from the normalization condition that the sum of the posterior over all possible models is unity.

To utilize this methodology, we must assign probabilities that accurately describe our state of knowledge. As a first pass, we begin with a Gaussian likelihood

$$P(D|M,I) = (2\pi\sigma^2)^{-N/2} Exp\left[-\sum_{\lambda}\frac{(F(\lambda)-D(\lambda))^2}{2\sigma^2}\right], \quad (6)$$

where $D(\lambda)$ represents the measured flux that constitutes our data, $F(\lambda)$ represents the modeled spectral flux described by (1), (2), and (3), and $\sigma$ is the expected squared deviation of the measurements from the true unknown values. The summation is over the $N$ values of measured $\lambda$, which can be indexed allowing for non-uniform sampling of the spectrum or accommodating missing data. Normally, $\sigma$ is unknown, and we can deal with this by integrating over all possible values of sigma. When using a Jeffrey's prior for $\sigma$, this results in the Student-t distribution [5, p. 54]

$$P(D|M,I) = \left[\sum_{\lambda}(F(\lambda)-D(\lambda))^2\right]^{-N/2}, \quad (7)$$

which conservatively accounts for the fact that the noise variance is unknown.

The majority of our prior information has gone into choosing the relevant model parameters and assigning a likelihood function that depends on the form of the spectrum model. Clearly, we can put limits on the possible cloud temperatures and PAH, Planck, and Gaussian contributions/amplitudes. For simplicity, the present results are obtained by assigning uniform priors over those ranges

$$P(A|I) = \frac{1}{A_{max} - A_{min}}. \quad (8)$$

However, future work will involve the use of more sophisticated priors. In the event that the average expected values of the amplitude $\bar{A}$ is known, an exponential prior is more appropriate [5, p. 117]

$$P(A|I) = \frac{1}{\bar{A}} \exp\left[-\frac{A}{\bar{A}}\right]. \quad (9)$$

Skilling also suggests a similar prior

$$P(A|I) = \frac{A_o}{(A_o + A)^2}, \quad (10)$$

which captures the expectation that the amplitude is on the order of $A_o$, but is more forgiving for occasional larger values. With the priors and the likelihoods defined, the next step is to explore to posterior probability and identify the high-probability solutions.

## 5. NESTED SAMPLING

To explore the posterior probability, we employ a new MCMC technique called Nested Sampling [5, 6]. Nested Sampling focuses on the joint probability

$$\begin{aligned}P(M,D|I) &= P(M|D,I)\,P(D|I) \\ &= P(D|M,I)\,P(M|I)\end{aligned} \quad (11)$$

which can be written as the posterior times the evidence, or equivalently, the likelihood times the prior. The algorithm maintains a set of samples distributed according to the prior probability, each with its associated likelihood. At each step, the sample with the least log likelihood is identified and used to set a likelihood threshold below which we no longer need to explore. That sample is discarded and replaced by duplicating and varying one of the other samples. The sample is varied so as to remain above the new likelihood threshold. This is performed by step-size monitoring and rejection of samples that cross the implicit threshold. All moves above the threshold are accepted. The result is a method that maintains an implicit boundary that iteratively contracts into the high probability regions (Figure 5).

One can use the discarded samples to numerically integrate the volume of the posterior, which enables one to compute the evidence. This will be extremely important in our problem since the evidence is critical in assessing the validity of the model order and the comparison across a variety of spectral models. From the list of discarded samples, one can compute mean quantities of the parameter values as well as the uncertainties in the estimates.

To implement model selection we have adopted Skilling's ON/OFF strategy [5, 6] that turns sources on and off in addition to varying their parameter values. The result is that we can compute the probability that a particular source is present in addition to its estimated contribution. This is implemented by introducing an additional binary parameter $\delta$, which modifies the PAH term in (2)

$$\sum_{p=1}^{P} c_p\,\delta_p\,PAH_p(\lambda) \quad (12)$$

so that $\delta_p$ turns PAH $p$ ON and OFF, and $c_p$ scales its contribution to the recorded spectrum. By separating the two effects, we can estimate both the probability that PAH $p$ is present and the degree to which it contributes to the spectrum.

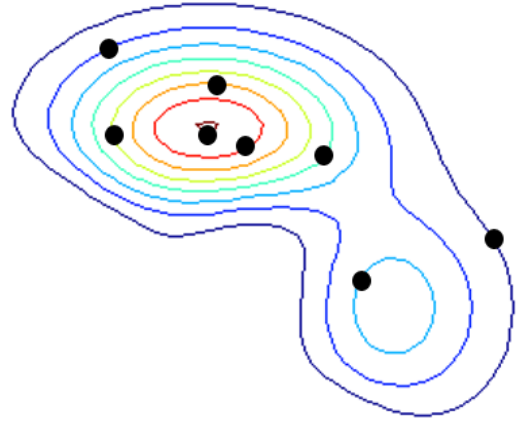

Figure 5. This cartoon shows a two-dimensional hypothesis space. Nested sampling relies on maintaining a set of samples distributed according to the prior probability. Associated with each sample in the space is a likelihood, which is used to define likelihood thresholds below which the algorithm will not explore. This implicitly defines boundaries which contract on the high probability regions.

This is handled in Nested Sampling by creating engines that perform various tasks with different rates that are set to be proportional to the prior probabilities. One such engine simply varies the present PAH contributions, another engine flips the state of a PAH from ON to OFF and vice versa, while yet another engine may swap one PAH for another with a similar spectrum. A wide variety of engines can be created that enable the algorithm to effectively explore the model parameter space.

## 6. RESULTS

At this stage in our research, we have implemented Nested Sampling for separation of the spectral background into *Planck* and *Gauss* spectra and have applied the algorithm to spectral data collected by the Infrared Space Observatory (ISO). In addition, we have separately worked with characterizing PAHs from artificial mixtures. The original expectation was to be able to estimate the background sources separately from the PAH contributions. However, due to the fact that there are both emission and absorption processes, this will lead to poor estimates making it necessary to perform a joint estimation.

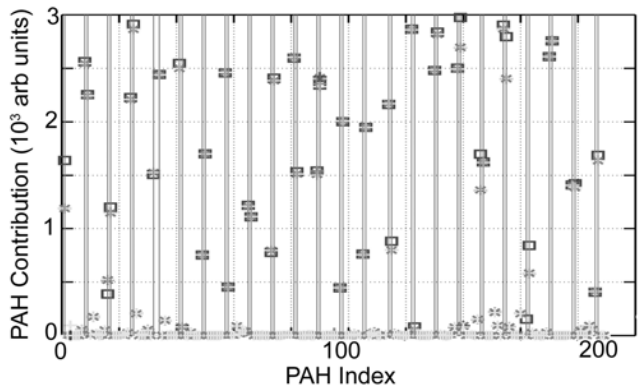

Figure 6. This shows the results of 7 early experiments aiming to estimate the contributions of 47 PAHs. The x-axis represents the PAH labels. The vertical lines indicate PAHs that were actually present in the synthetic data set. The squares represent their true contributions, and the stars the estimates. Most present PAHs have been detected. The ON-OFF technique is not being applied here, so all PAHs are estimated to have some, albeit very small, contribution.

Figure 3 shows a spectrum taken from the Orion Bar in the Orion nebula, illustrated in Figure 1. The black curve is the original data, and the gray curve is the background estimation obtained from nested sampling. There is one blackbody radiator at a temperature of $61.043 \pm 0.004$ K, and possibly a second (36.3% chance), at a temperature around 18.8 K.

In Figure 6 we show the results of seven experiments on data consisting of a synthetic mixture of 47 PAHs. This example is essentially, a non-negative least squares fit and demonstrates feasibility (see caption). We are currently working to apply the Nested Sampling ON-OFF procedure to the PAH mixtures.

## 7. CONCLUSIONS

As we described in the Introduction, this is clearly a formidable problem and the probability of our final success is not clear. However, without the ability to traverse light-years of distance and collect physical samples, a single infra-red spectrum of an astronomical site is all that is available for study. A careful study of the successes and failures of our signal processing techniques applied to this problem will be critical to understand what additional information, if any, about the interstellar environment and the underlying principles of astrochemistry is necessary to obtain the most complete understanding of these important processes.

Our future work will consist of making the models of the spectral components more physically accurate, increasing the number of PAH species considered, and identifying relevant classes of PAHs in the event that individual PAH species are simply not identifiable.

## 8. REFERENCES

[1] L.J. Allamandola, A.G.G.M. Tielens, J.R. Barker, "Polycyclic aromatic hydrocarbons and the unidentified infrared emission bands: Auto exhaust along the Milky Way!" *Astrophys. J. Letters,* 290, L25, 1985.

[2] L.J. Allamandola, D.M. Hudgins, S.A. Sandford, "Modeling the unidentified infrared emission with combinations of polycyclic aromatic hydrocarbons," *ApJ*, 511, L115-119, 1999.

[3] K.H. Knuth, "Informed source separation: A Bayesian tutorial," In: B. Sankur, E. Çetin, M. Tekalp, E. Kuruoğlu (eds.), Proceedings of the 13th European Signal Processing Conference (EUSIPCO 2005), Antalya, Turkey, 2005.

[4] E. Peeters, S. Hony, C. van Kerckhoven, A.G.G.M. Tielens, L.J. Allamandola, D.M. Hudgins, and C.W. Bauschlicher, "The rich 6 to 9 vec mu m spectrum of interstellar PAHs," *A&A*, 390, 1089-1113, 2002.

[5] D.S. Sivia, J. Skilling "Data Analysis: A Bayesian Tutorial", 2nd Ed. Oxford University Press, Oxford, 2006.

[6] J. Skilling, "Turning ON and OFF," In: K.H. Knuth, A. Abbas, R. Morris (eds.), Bayesian Inference and Maximum Entropy Methods in Science and Engineering, San José, California, USA, AIP Conference Proceedings 803, American Institute of Physics, Melville NY, pp. 3-24.